\documentclass[referee]{raa}            

\usepackage{caption,subcaption}
\usepackage{graphicx,times}          
\usepackage{natbib}
\usepackage{overpic}
\graphicspath{{./}{figures/}}
\usepackage{amssymb,amsmath}
\bibpunct{(}{)}{;}{a}{}{,}
\usepackage{hyperref}
\hypersetup{colorlinks,				
	linkcolor=blue,
	citecolor=blue}

\begin{document}

   \title{Investigating the energy distribution of the high-energy particles in the Crab nebula
}

   \volnopage{Vol.0 (20xx) No.0, 000--000}      
   \setcounter{page}{1}          

   \author{Lu\ Wen
      \inst{1}
   \and Ke-Yao\ Wu
      \inst{1}
   \and Huan\ Yu
      \inst{2}
   \and Jun\ Fang
      \inst{1,3}
   }

    \institute{Key Laboratory of Astroparticle Physics of Yunnan Province, Yunnan University, Kunming 650091, China; \\
        \and
             Department of Physical Science and Technology, Kunming University, Kunming 650214, China; {\it yuhuan.0723@163.com}\\
        \and
            Department of Astronomy, Yunnan University, Kunming 650091, China; {\it fangjun@ynu.edu.cn}
\vs\no
   {\small Received~~20xx month day; accepted~~20xx~~month day}}

\abstract{
The Crab nebula is a prominent pulsar wind nebula (PWN) detected in multiband observations ranging from radio to very high-energy (VHE) $\gamma$-rays. Recently, $\gamma$-rays with energies above $1 \mathrm{PeV}$ had been detected by the Large High Altitude Air Shower Observatory (LHAASO), and the energy of the most energetic particles in the nebula can be constrained. In this paper, we investigate the broadest spectral energy distribution of the Crab nebula and the energy distribution of the electrons emitting the multiwavelength nonthermal emission based on a one-zone time-dependent model. The nebula is powered by the pulsar, and high-energy electrons/positrons with a broken power-law spectrum are continually injected in the nebula as the pulsar spins down. Multiwavelength nonthermal emission is generated by the leptons through synchrotron radiation and inverse Compton scattering. Using appropriate parameters, the detected fluxes for the nebula can be well reproduced, especially for the $\gamma$-rays from $10^2\,\mathrm{MeV}$ to $1\,\mathrm{PeV}$. The results show that the detected $\gamma$-rays can be produced by the leptons via the inverse Compton scattering, and the lower limit of the Lorentz factor of the most energetic leptons is $\sim 8.5\times10^{9}$. It can be concluded that there are electrons/positrons with energies higher than $4.3$\,PeV in the Crab nebula.
}

   \authorrunning{Wen et al.}            
   \titlerunning{The distribution of particles in the Crab nebula}  

   \maketitle

\section{Introduction}
\label{intro}
The Crab nebula is powered by an energetic pulsar PSR J0534$+$2200 which is generated from a core-collapse supernova in 1054 AD recorded in Chinese astronomers \citep{SL21}, and the PWN has been detected over a wide range of wavelengths from radio to PeV $\gamma$-rays \citep[e.g.][]{AG03,Mea10,LT19,AA21,LC21}. It is currently a young PWN with no apparent shell structure on the outside because the remnant has not yet interacted with enough of the surrounding medium to observe a supernova shell \citep{SG06}. The nonthermal emission of the nebula spans 20 magnitudes from radio ($10^{-5} \, \mathrm{eV}$)  to $\geq 100 \, \mathrm{TeV}$, and it is now the standard candle for calibration of various detectors \citep{AA06,AB08}.

Particles accelerated in the Crab nebula have energies up to $\sim1$\,PeV based on the results of the detected high-energy emission associated with it.
Analysis of X-ray emission with Chandra for the nebula shows that there exists electrons with  energies above $100 \, \mathrm{TeV}$ \citep{WJ00}; moreover, $\gamma$-rays with energy $\sim 80 \, \mathrm{TeV}$ indicate the particles having energies up to $\sim 10^{3} \, \mathrm{TeV}$ in the PWN \citep{AA04,AA06}. Recently, HAWC \citep{AA17,AA19} and Tibet AS+MD \citep{AB19} have detected $\gamma$-rays from the Crab nebula with energies above $100 \, \mathrm{TeV}$, and $\gamma$-ray photons with energies up to $\sim1 \, \mathrm{PeV}$ from it have been collected with LHAASO \citep{LC21}. The observed fluxes around $100 \, \mathrm{TeV}$ are consistent with a smooth extrapolation of the lower-energy spectrum.

The radiative mechanisms involved in producing the multiband nonthermal emission from the Crab nebula have been widely studied. \citep{AA96} proposed that the spectrum from radio to $\mathrm{MeV}$ energy band is produced by synchrotron radiation from electrons, and that $\gamma$-rays in the higher energy band are produced by inverse Compton scattering from soft photons, while at the same time bremsstrahlung may make a contribution to the production of high energy $\gamma$-rays.  The ion acceleration in PWNe had been studied, and it was concluded that PWNe could be the source of cosmic rays \citep{FK13,CB14,KA15,GC20}. \citet{ZC20} constrained the contribution/impact of the hadronic process on high-energy $\gamma$-ray emission for the Crab nebula.

In this paper, motivated by the recent detections with energies up to $\sim 1 \, \mathrm{PeV}$ $\gamma$-rays, we use a one-zone time-dependent model for the multiband nonthermal emission from a PWN to study the energy distribution of the high-energy particles in the Crab nebula. In Section \ref{model}, we briefly describe the one-zone time-dependent model. In Section \ref{results}, we use the model to investigate the multiband radiative properties of the Crab nebula, and the parameters can be constrained by comparing the resulting SED with the detected fluxes. In Section \ref{summary}, the discussion and summary are indicated.

\section{THE MODEL FOR THE MULTIBAND NONTHERMAL EMISSION FROM PWNE}
\label{model}
A termination shock can be produced as the relativistic winds from a pulsar interacting with the ambient medium. High-energy leptons (electrons/positrons) are accelerated by the shock, and they are  continually injected in the nebula with a rate of $Q(\gamma, t)$. The distribution of the high-energy particles in the nebula evolves based on the equation \citep{FZ10,MT12}
\begin{equation}
\frac{\partial N(\gamma, t)}{\partial t} = \frac{\partial}{\partial\gamma}[\dot{\gamma}(\gamma, t)N(\gamma, t)] - \frac{N(\gamma, t)}{\tau(\gamma, t)} + Q(\gamma, t) \;,
\label{eq:ngammat}
\end{equation}
where $N(\gamma, t)$ is the number of the particles at time $t$ with lorentz factors between $\gamma$ and $\gamma + d\gamma$. The particles generate multiband nonthermal emission through synchrotron radiation and inverse Compton scattering, and $\dot{\gamma}(\gamma, t)$ is the energy loss rate of the particles with a lorentz factor of $\gamma$. These high-energy leptons can escape from the nebula due to Bohm diffusion, and $\tau(\gamma, t)$ represents the escape time.
For the Crab nebula, it is usually assumed that the particles are injected with a broken power-law spectrum, i.e.,
\begin{equation}
Q(\gamma, t) = Q_0(t) \left\{
  \begin{array}{cc}
    \left( \frac{\gamma}{\gamma_{\mathrm{b}}} \right)^{-\alpha_1} & \mathrm{if~} \gamma \leq \gamma_{\mathrm{b}}\;, \\
    \left( \frac{\gamma}{\gamma_{\mathrm{b}}} \right)^{-\alpha_2} & \mathrm{if~} \gamma_{\mathrm{b}} < \gamma \leq \gamma_{\mathrm{max}}\;,
    \label{eq:qgamma2}
   \end{array}
\right.
\end{equation}
\begin{equation}
\gamma_{\mathrm{max}}(t) = \frac{\varepsilon e \kappa}{m_{\mathrm{e}}c^2}\left ( \frac{\eta L(t)}{c} \right )^{1/2},
\label{eq:gmax}
\end{equation}
where the parameters $\alpha_1$ and $\alpha_2$ are the spectral indices for the particles with $\gamma$ below and above the break Lorentz factor $\gamma_{\mathrm{b}}$, respectively, and $\eta$ is the magnetic energy fraction, $\varepsilon$ is the fractional size of the
radius of the shock. $m_e$, $e$, $c$ are the electron mass, electron charge and speed of light, respectively. As in \citet{MT12}, the magnetic compression ratio $\kappa$ is adopt to be $3$ in this paper.

The rotation period ($P$) of the pulsar increases gradually with a period-derivative of $\dot{P}$ as it spins down, and the spin-down luminosity can be derived with
\begin{equation}
L(t) = 4\pi^2 I \frac{\dot{P}}{{P}^3} = L_0 \left ( 1+\frac{t}{\tau_0}\right )^{-\frac{n+1}{n-1}},
\label{eq:lt}
\end{equation}
where the initial spin-down time-scale of the pulsar is
\begin{equation}
\tau_0 = \frac{2\tau_c}{n-1} - t_{\mathrm{age}},
\label{eq:tau0}
\end{equation}
where $I=10^{45}\mathrm{g}\,\mathrm{cm}^{2}$ is the moment of inertia of the pulsar, $L_0$ is the initial luminosity, $n$ is the braking index of the pulsar with an age of $t_{\mathrm{age}}$, and $\tau_c = P/2\dot{P}$ is characteristic age.

Assuming a fraction of the spin-down luminosity is transferred to the magnetic field which is homogeneous in the nebula, the magnetic field can be obtained with \citep{MT12}
\begin{equation}
B(t) = \sqrt{ \frac{3(n-1)\eta L_0 \tau_0}{R^3_{\mathrm{PWN}}(t)} \left [1-\left( 1+\frac{t}{\tau_0}\right )^{-\frac{2}{n-1}}\right ] },
\label{eq:bt}
\end{equation}
where $R_{\mathrm{PWN}}$ is the radius of the nebula at time $t$. In the expanding phase, the nebula expands according to \citep{VA01}
\begin{equation}
R_{\mathrm{PWN}}(t) = 0.84 \left( \frac{L_0 t}{E_0} \right)^{1/5} \left( \frac{10E_0}{3M_{\mathrm{ej}}} \right )^{1/2}t,
\label{eq:rpwn}
\end{equation}
where $E_0 =10^{51}\mathrm{erg}$ and $M_{\mathrm{ej}}=9.5M_{\odot}$ are the kinetic energy  and the mass of the supernova ejecta, respectively. Assuming all of the spin-down luminosity except the fraction transferred into the magnetic field are used to accelerate the injected particles, the normalized factor $Q_0(t)$ can be derived from
\begin{equation}
\int \gamma m_e c^2 Q(\gamma, t)d\gamma = (1-\eta)L(t).
\label{eq:q0}
\end{equation}

\begin{figure}
        \centering
        \includegraphics[width=0.6\textwidth]{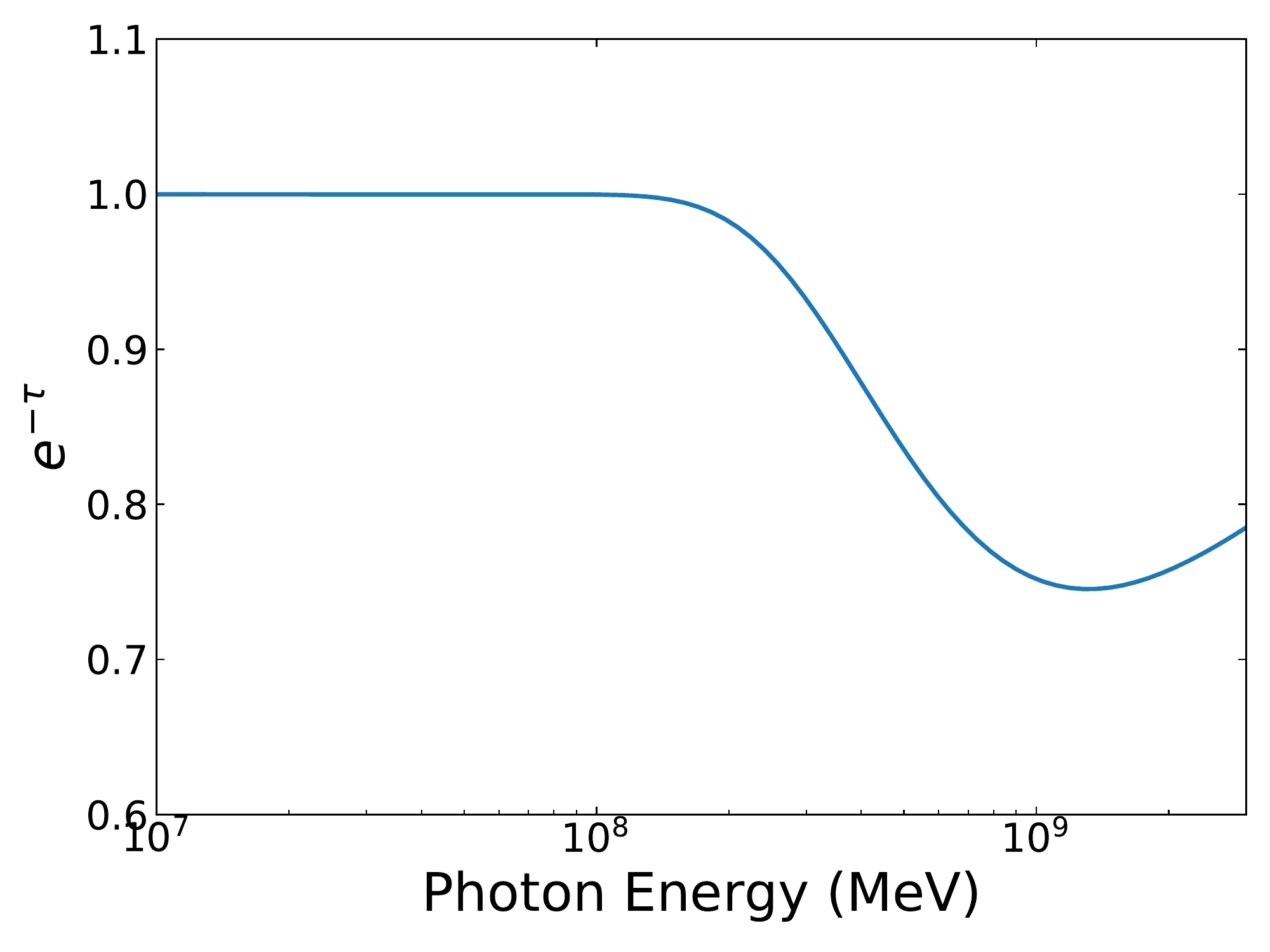}
        \caption{$\gamma$-ray opacity of the Crab nebula for the pair production via the interaction of the $\gamma$-rays with the CMB photons.}
        \label{fig:opa}
\end{figure}

VHE $\gamma$-ray photons in the Galaxy can interact with those from the CMB and the interstellar
radiation fields (ISRF), so the $\gamma$-rays can be attenuated due to the process of pair ($e^{\pm}$) production. Especially, in the Galactic center,
the attenuation of $\gamma$-rays with energy above $100$\,TeV is significant due to the high energy density of the ISRF \citep{Mea06}. The cross section for the pair production due to the interaction of the two photons with energies $\varepsilon_1$ and $\varepsilon_2$ is \citep[e.g.,][]{L06}
\begin{equation}
\sigma = \frac{\pi r_e^2}{2}(1-\beta^2)\left [2\beta(\beta^2-2) + (3 - \beta^4)\ln\left(\frac{1+\beta}{1-\beta}\right)\right ],
\label{eq:segma}
\end{equation}
where
\begin{equation}
\beta = \left [ 1 - \frac{m_e^2 c^4}{\varepsilon_1 \varepsilon_2}  \right ]^{\frac{1}{2}},
\label{eq:beta}
\end{equation}
$r_e$ is the classical electron radius.
For the Crab nebula with a distance of $2\,$kpc, the attenuation of the VHE $\gamma$-rays is mainly due to the interaction with the CMB photons, and the minimum opacity ($e^{-\tau}$, $\tau$ is the optical depth) is $\sim 0.75$ at $\sim 1$\,PeV \citep{Cea21}. In this paper, we take into account the attenuation of the VHE $\gamma$-rays which interacts with the CMB photons to product pairs.

\section{Results}
\label{results}

\begin{figure}
        \centering
        \includegraphics[width=0.6\textwidth]{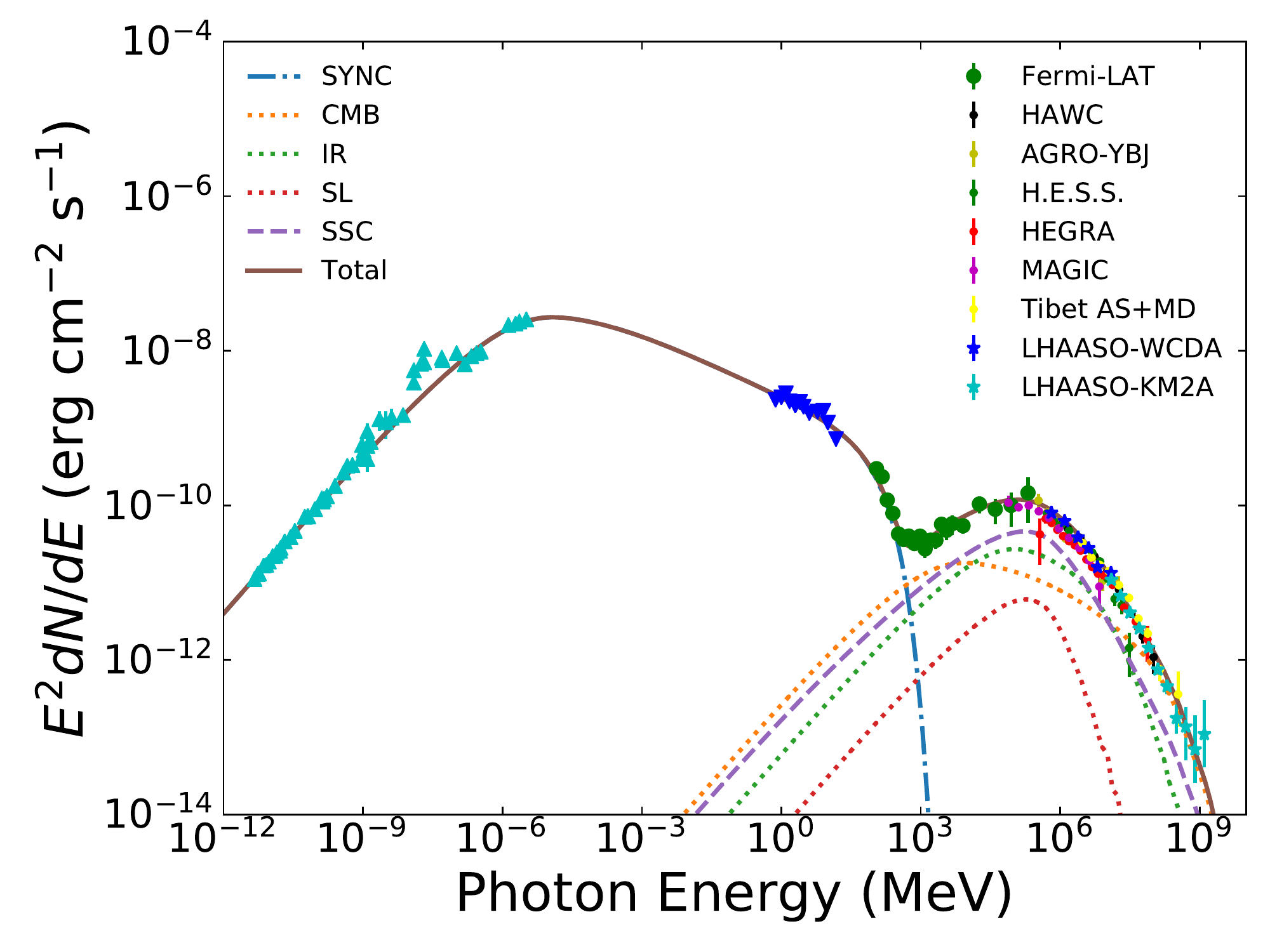}
        \caption{Comparison of the resulting SED of the Crab nebula, which includes synchrotron radiation, inverse Compton scattering off CMB, FIR, starlight and synchrotron photons, with the multiband detected fluxes.The observed fluxes of radio \citep{MM10}, infrared \citep{GT04,TG06}, optical \citep{VW93}, X-ray \citep{KH01} and $\gamma$-ray bands with Fermi-LAT \citep{BS12}, HAWC \citep{AA19}, ARGO-YBJ \citep{BB15}, H.E.S.S. \citep{AA06}, HEGRA \citep{AA04}, MAGIC \citep{AA15}, Tibet AS+MD \citep{AB19}, LHAASO-WCDA, and LHAASO-KM2A \citep{LC21} are shown in the figure for comparison.}
        \label{fig:crab}
\end{figure}

\begin{figure}
        \centering
        \includegraphics[width=0.5\textwidth]{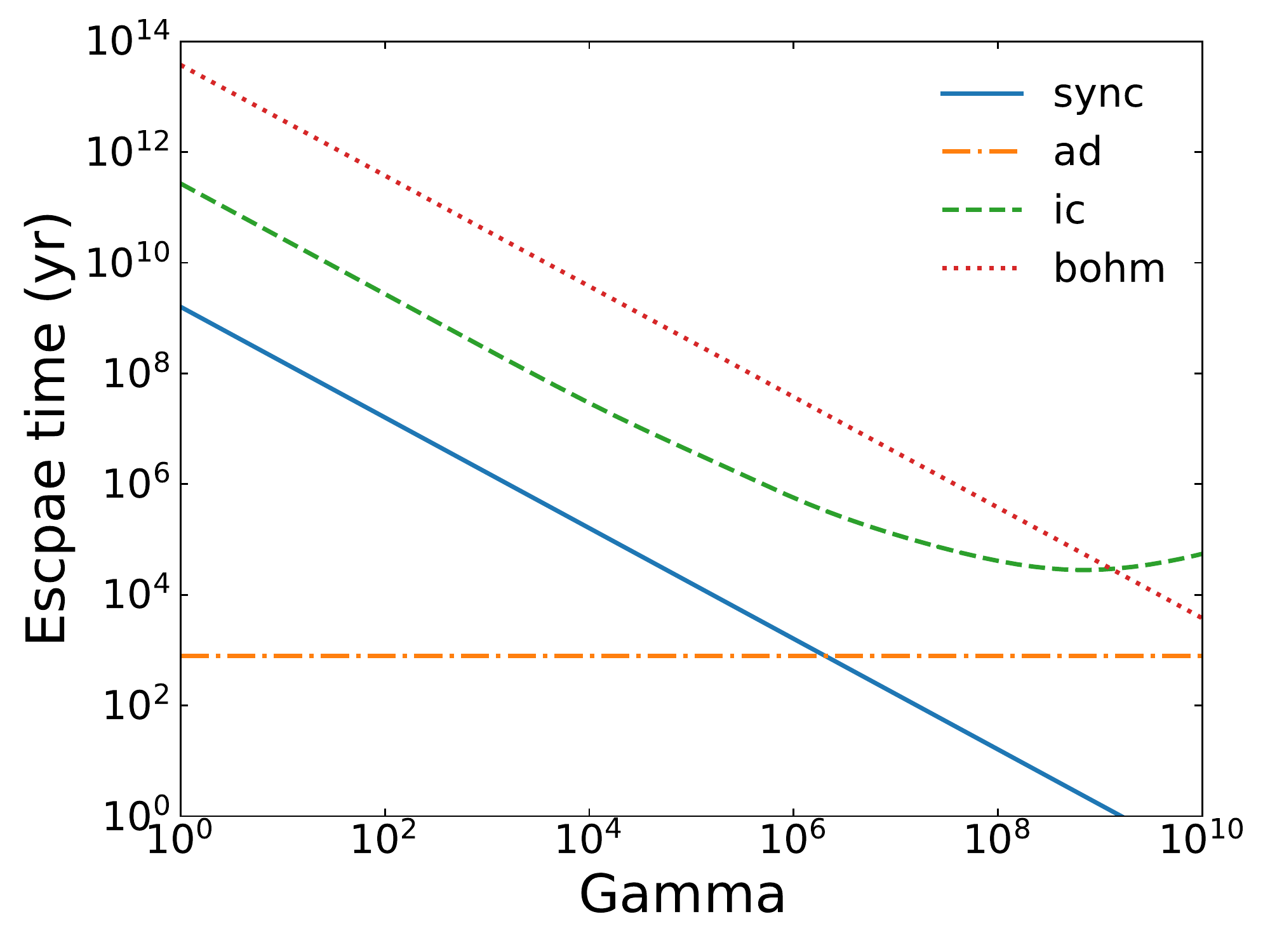}\includegraphics[width=0.5\textwidth]{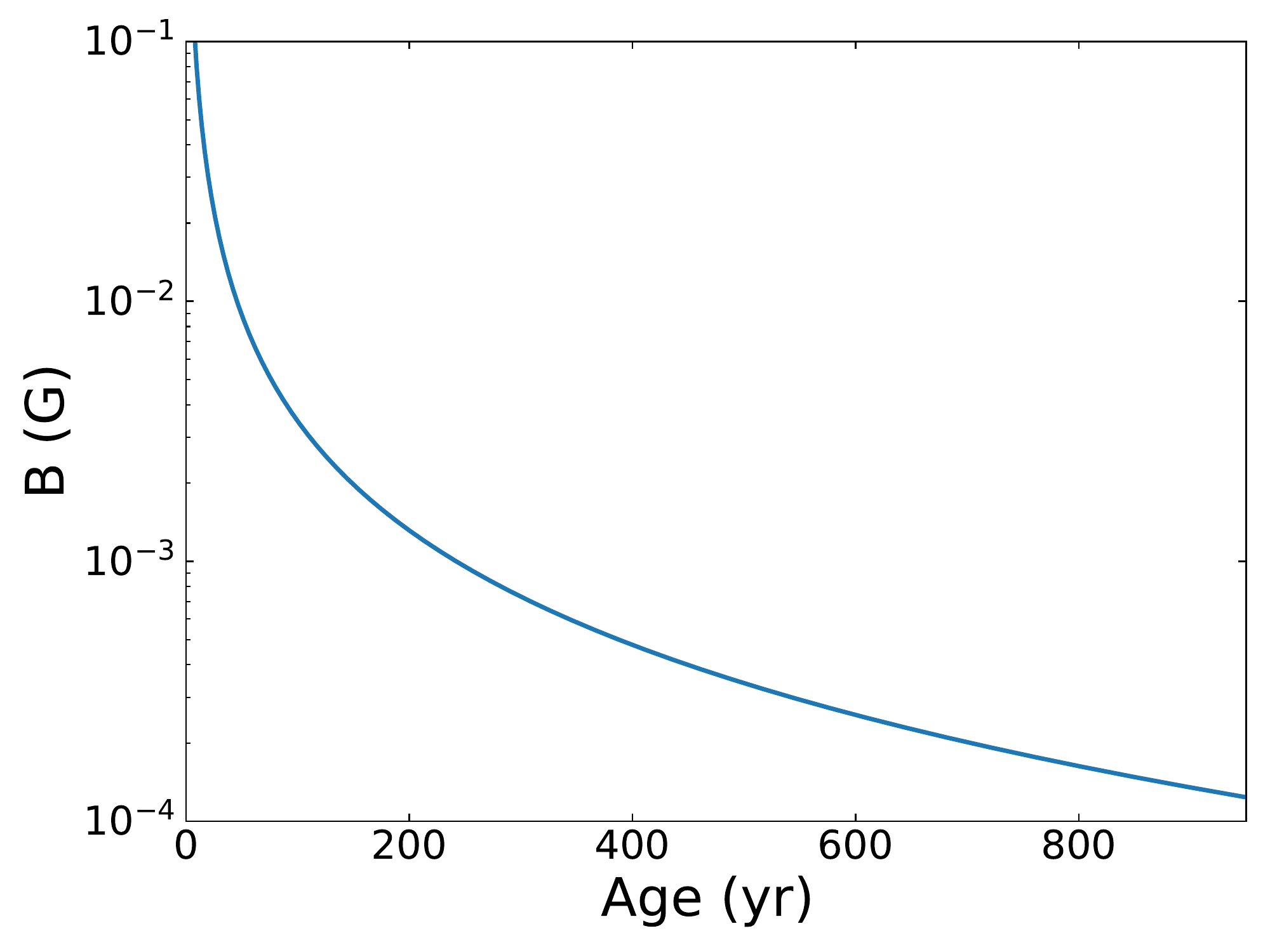}
        \caption{Left panel: cooling times for the synchrotron radiation (solid line), the adiabatic loss (dash-dotted line), the inverse Compton scattering (dashed line), respectively, and the escape time for the Bohm diffusion (dotted line) with $\varepsilon=0.28$ at $950 \, \mathrm{yr}$. Right panel: the magnetic field strength over time with $\varepsilon=0.28$. }
        \label{fig:escapetime}
\end{figure}

The Crab nebula is powered by the pulsar with $P = 33.4 \, \mathrm{ms}$, $\dot{P} = 4.21 \times 10^{-13} \, \mathrm{s \, s^{-1}}$, $T_{\rm age} = 950$\,yr at a distance of $d=2 \, \mathrm{kpc}$  \citep{TM93}, and the braking index is $n =2.509$ \citep{LP88}. Following \citet{TC14}, the four different seed photon fields involved in the inverse Compton scattering in the Crab nebula are the CMB radiation with a temperature of $T_{\mathrm{CMB}}=2.73\, \mathrm{K}$ and an energy density of $\mathrm{U_{CMB}} = 0.25 \, \mathrm{eV\,cm}^{-3}$, the FIR radiation with $T_{\mathrm{FIR}}=70\, \mathrm{K}$ and $\mathrm{U_{FIR}} = 0.5 \, \mathrm{eV\,cm}^{-3}$, the starlight radiation with $T_{\mathrm{SL}}=5000\, \mathrm{K}$ and $\mathrm{U_{FIR}} = 1.0 \, \mathrm{eV\,cm}^{-3}$, and the synchrotron radiation in the nebula.

Assuming the magnetic energy fraction is $\eta = 0.02$ and the spectrum of the injected electrons/positrons is a broken power-law with $\alpha_1 = 1.61$, $\alpha_2 = 2.56$, $\gamma_{\mathrm{b}} = 2\times10^6$,
$\varepsilon = 0.28$, the resulting SED of the Crab nebula at $T_{\rm age} = 950 \, \mathrm{yr}$ is shown in Fig. \ref{fig:crab}.
The radius of the nebula and the maximum energy of the leptons from the model are about $2.13 \, \mathrm{pc}$ and $4.3 \, \mathrm{PeV}$, respectively. The detected IR bump at $\sim 0.01 \, \mathrm{eV}$ is thought to be thermal dust emission of the PWN \citep{ZF15}. The detected fluxes of the multiband nonthermal emission, especially in the $\gamma$-rays with energies above $10^3 \, \mathrm{MeV}$, can be well reproduced with the model. As indicated by the Fig\ref{fig:crab}, the resulting $\gamma$-ray spectrum from the model is well consistent with the latest LHAASO results.
Data points with LHAASO-KM2A in the energy range $12 - 1300 \, \mathrm{TeV}$ are indicated by cyan asterisks, and those with LHAASO-WCDA in the energy range $0.65 - 12.36 \, \mathrm{TeV}$ are indicated by blue asterisks \citep{LC21}.
The lower-energy nonthermal emission from radio to $1 \, \mathrm{GeV}$ comes from synchrotron radiation by the high-energy electrons/positrons. The higher-energy component of the nonthermal emission with energies above above $\sim 1 \, \mathrm{GeV}$ is produced via the inverse Compton scattering.

At $T_{\rm age} = 950 \, \mathrm{yr}$, the magnetic filed strength in the nebula is $B = 123.6 \, \mathrm{\mu G}$, and the cooling of the particles with $\gamma<10^6$ is mainly determined by the adiabatic loss due to the expansion of the nebula. However, the particles with higher energies encounter strong synchrotron radiation, and the synchrotron radiation is the dominate process in cooling the particles.
As illustrated in the left panel of Fig. \ref{fig:escapetime}, with $\gamma>10^6$, the cooling of the particles is mainly determined by the synchrotron radiation, which results in the particle spectrum becomes softer at higher energies with $\gamma>10^6$.

\begin{figure}
        \centering
        \includegraphics[width=0.6\textwidth]{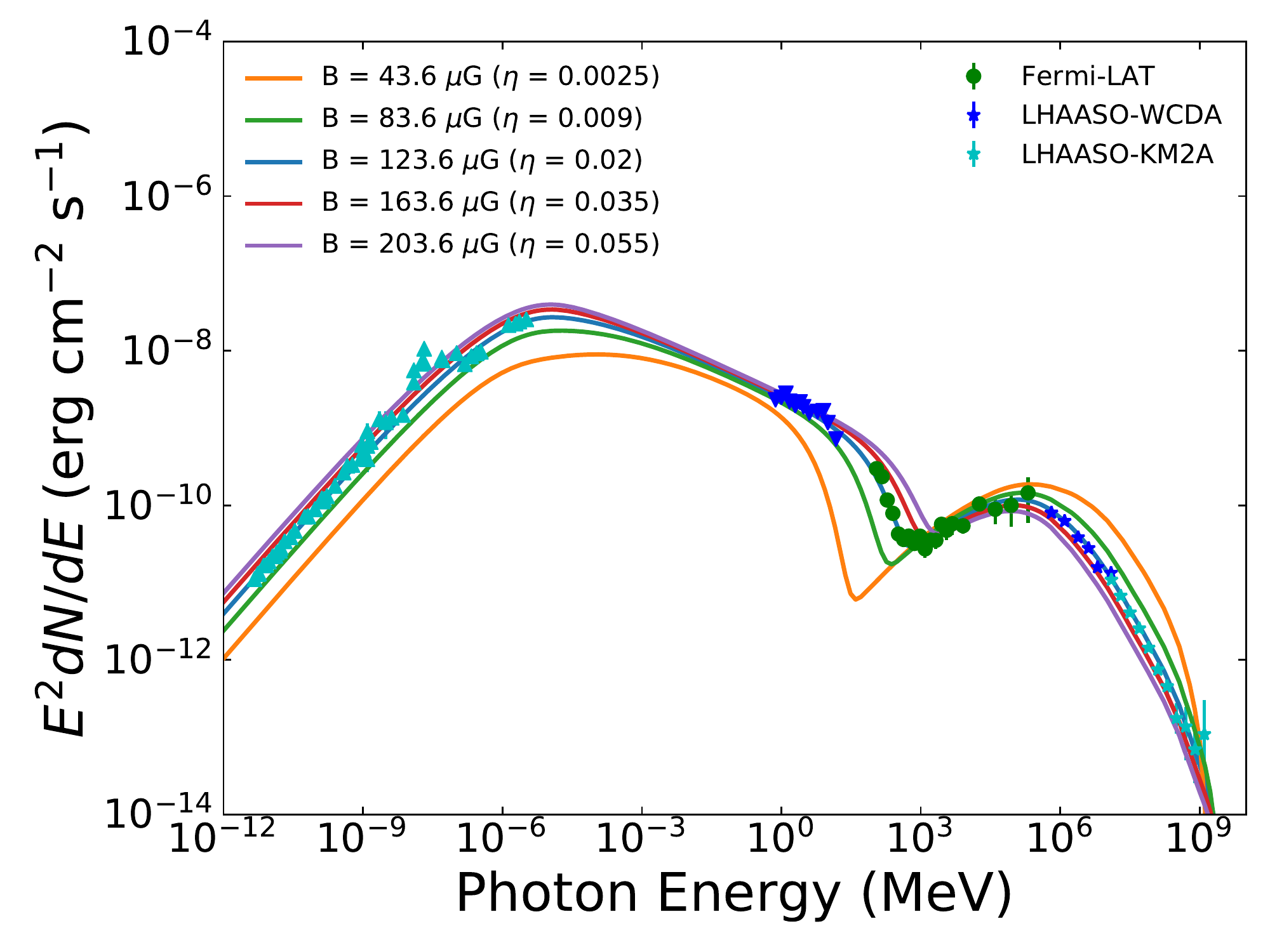}
        \caption{The resulting SEDs with different magnetic field strength (magnetic energy fraction). The references on the observations are the same as Fig. \ref{fig:crab}.}
        \label{fig:bt_yanhua}
\end{figure}

\begin{figure}
        \centering
        \includegraphics[width=0.5\textwidth]{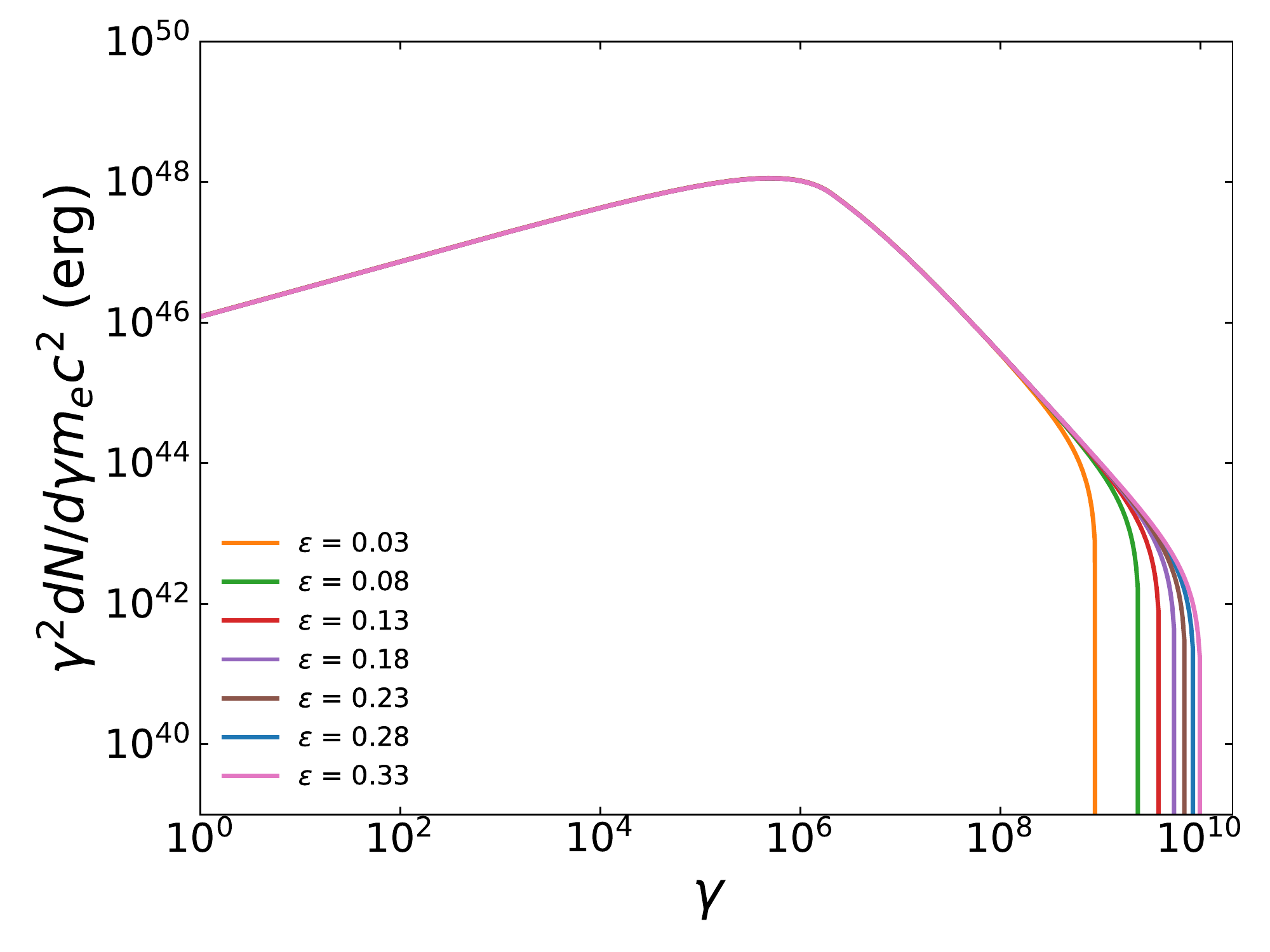}\includegraphics[width=0.5\textwidth]{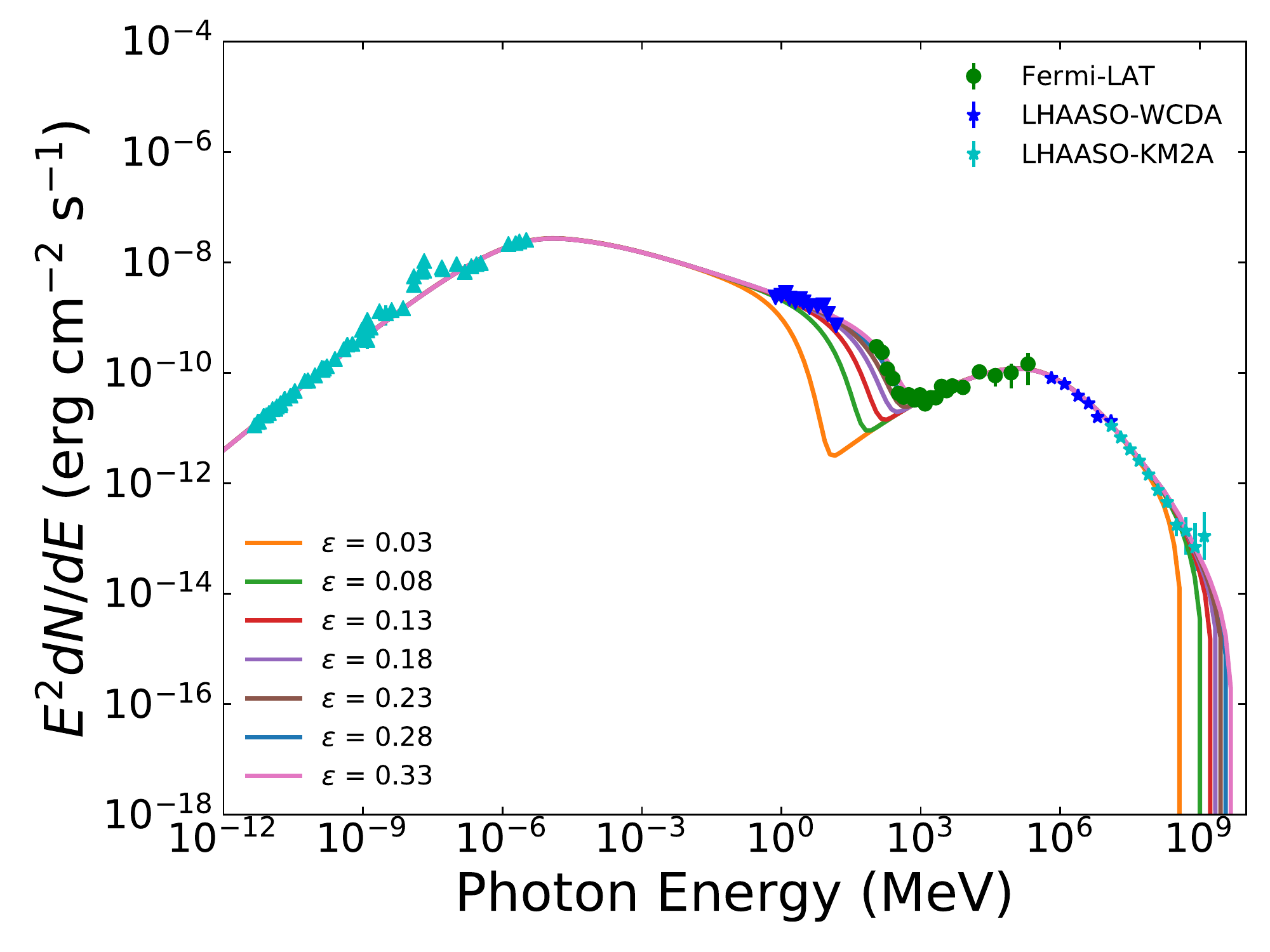}
        \caption{Left panel: the particle spectra of the electrons/positrons with different values of $\varepsilon$ at $T_{\rm age}$. Right panel: the SEDs for each value of $\varepsilon$.}
        \label{fig:crab_epsilon}
\end{figure}

The magnetic field in the PWN diminishes gradually due to the expansion of the nebula (the right panel of Fig. \ref{fig:escapetime}). The magnetic field strength of the nebula is $123.6 \, \mathrm{\mu G}$ at $950 \, \mathrm{yr}$ with $\eta=0.28$. The influence of the magnetic field (the magnetic energy fraction $\eta$) on the SED at $950 \, \mathrm{yr}$ is shown in Fig. \ref{fig:bt_yanhua}. The SEDs with $B = 43.6 \, \mathrm{\mu G}$ ($\eta$ = 0.0025), $B = 83.6 \, \mathrm{\mu G}$ ($\eta$ = 0.009), $B = 123.6 \, \mathrm{\mu G}$ ($\eta$ = 0.02), $B = 163.6 \, \mathrm{\mu G}$ ($\eta$ = 0.035) and $B = 203.6 \, \mathrm{\mu G}$ ($\eta$ = 0.055) are indicated with different colors,  respectively. A weaker magnetic field results in a lower radio flux for the synchrotron radiation. However, with a lower magnetic field strength, the break energy of the particle spectrum due to the cooling of the synchrotron radiation and the cutoff energy of the resulting photon spectrum in the $\mathrm{TeV}$ band are higher. When $B > 123.6 \, \mathrm{\mu G}$, the fluxes above $1 \, \mathrm{TeV}$ are lower than those detected with LHAASO-WCDA and LHAASO-KM2A.

Fig. \ref{fig:crab_epsilon} shows the particle spectra of the electrons/positrons and the resulting SEDs with different values of $\varepsilon$. We can see that as the $\varepsilon$ gets larger, the maximum Lorentz factor of the particles also increases significantly. The $\gamma$-ray fluxes observed by the Fermi-LAT can be used to constrain the lower limit of the maximum energy of the electrons/positrons. From the right panel of Fig. \ref{fig:crab_epsilon}, the LHAASO-KM2A data can be well explained if $\varepsilon > 0.08$. With $\varepsilon \leq 0.23$, the resulting fluxes ranging from $10^2 - 10^3$\,MeV are lower than those detected with Fermi-LAT.
With $\varepsilon = 0.28$, it corresponds to a maximum Lorentz factor of $\sim 8.5\times10^{9}$ and a maximum energy of $\sim 4.3 \, \mathrm{PeV}$.

\section{Summary and discussion}
\label{summary}

With a one-zone time-dependent model for the multiband nonthermal emission from PWNe, we have studied the nonthermal radiative properties of the Crab nebula. With appropriate parameters, i.e., $\varepsilon = 0.28$, $\eta = 0.02$, $\alpha_1 = 1.61$, $\alpha_2 = 2.56$, $\gamma_{\mathrm{b}} = 2\times10^6$, the model can reproduce the detected fluxes of the Crab nebula from radio to VHE $\gamma$-ray band.
We assume the particle energy spectrum is a broken power-law, which has been used extensively in other papers \citep[e.g.][]{TC14}. At an age of $950 \, \mathrm{yr}$, the calculated PWN radius is $2.13 \, \mathrm{pc}$, the magnetic field strength of the nebula is $123.6 \, \mathrm{\mu G}$ and the maximum Lorentz factor is $\sim 8.5\times10^{9}$ (the maximum energy of the injected leptons is $\sim 4.3 \, \mathrm{PeV}$).

The origin of the electrons/positrons involved in producing the multiband nonthermal emission of the Crab nebula is still under debating. The Crab nebula is currently a young PWN with no apparent shell structure on the outside because the remnant has not yet interacted with enough of the surrounding medium to observe a supernova shell \citep{SG06}. The termination shock, which is generated by the interaction of the relativistic wind from the pulsar with the surrounding medium,  can accelerate particles to energies of several hundred $\mathrm{TeV}$ or more, and these accelerated particles interact with the surrounding magnetic field, soft photons and interstellar matter, emitting radiation ranging from radio and X-rays to VHE $\gamma$-rays. \citet{AA96} assumed that the electrons within the Crab nebula contain both radio and wind electrons, with the wind electrons producing high-energy $\gamma$-rays from the nebula.

The Crab nebula is one of the prominent and widely studied VHE $\gamma$-ray sources.
The model in this paper shows that the detected $\gamma$-rays are from the inverse Compton scattering of the electrons/positrons which also emit the nonthermal emission from radio to X-rays, and it predicts the fluxes above $1 \, \mathrm{PeV}$ for the Crab nebula. Further detections for the Crab nebula at PeV $\gamma$-rays are important for determine whether there is another component of hadronic particles in the nebula to produce $\gamma$-rays.

\noindent {\bf Acknowledgments}

This work is supported by NSFC grants under nos. 11873042, U2031107, and 12063004,
the Program for Excellent Young Talents, Yunnan University (WX069051, 2017YDYQ01),
the Candidate Talents Training Fund of Yunnan Province (2017HB003),
the National Key R\&D Program of China (2018YFA0404204).

\end{document}